\documentclass[oldversion,letter]{aax}

\usepackage{txfonts}
\usepackage{graphicx}
\usepackage{natbib}

\bibpunct{(}{)}{;}{a}{}{,} 

\graphicspath{{figs/}}

\def\msun{M$_\odot$}

\def\xmmu1007{XMMU\,J1007}
\def\kmps{km s$^{-1}$}

\begin{document}
        \title{XMMU\,J100750.5+125818: A strong lensing cluster at
        $z=1.082$\thanks{Based on observations made with ESO Telescopes at the
        La Silla or Paranal Observatories under programmes 78.A-0265 and
        80.A-0659}\fnmsep\thanks{Based on observations obtained with XMM-Newton, an
        ESA science mission with instruments and contributions directly funded
        by ESA Member States and NASA}}  

        \author{A.D. Schwope\inst{1}
                \and G. Lamer\inst{1}
                \and A.~de Hoon\inst{1}
                \and J. Kohnert\inst{1}
                \and H.~B\"ohringer\inst{2}
                \and J.P.~Dietrich\inst{3}
                \and R. Fassbender\inst{2}
                \and J.~Mohr\inst{4,2,5}
                \and M.~M\"{u}hlegger\inst{2}
                \and D.~Pierini\inst{2}
                \and G.W.~Pratt\inst{6,2}
                \and H.~Quintana\inst{7}
                \and P.~Rosati\inst{8}
                \and J.~Santos\inst{9}
                \and R.~\v{S}uhada\inst{2}
        }

        \offprints{A.~Schwope, \email{aschwope@aip.de}}

        \institute{Astrophysikalisches Institut Potsdam, An der Sternwarte 16,
                14482 Potsdam, Germany 
                \and 
Max-Planck-Institut f\"ur Extraterrestrische Physik,
                85748 Garching, Germany 
                \and 
Physics Department and  Michigan Center for Theoretical Physics, 
University of Michigan, 450 Church St, Ann Arbor, MI 48109-1040, USA
\and
Department of Physics, Ludwig-Maximilians-Universit\"{a}t,
                Scheinerstra\ss e 1, 81679 Munich, Germany 
                \and 
Excellence Cluster Universe, Boltzmannstr.\ 2, 85748 Garching, Germany
\and
Laboratoire AIM, IRFU/Service d'Astrophysique - CEA/DSM -
                CNRS - Universit\'{e} Paris Diderot, B\^{a}t. 709, CEA-Saclay,
                F-91191 Gif-sur-Yvette Cedex, France 
              \and 
Departamento de Astronomia y Astrofisica, Pontificia
                Universidad Catolica de Chile, Casilla 306, Santiago, 22,
                Chile 
                \and 
European Southern Observatory, Karl-Schwarzschild-Strasse
                2, 85748 Garching, Germany
              \and 
INAF-Osservatorio Astronomico di Trieste, via Tiepolo 11,
                34131 Trieste, Italy 
} 

\date{Received / Accepted }

\abstract{We report on the discovery of the X-ray luminous cluster
  XMMU\,J100750.5+125818 at redshift 1.082 based on 19 spectroscopic members, 
  which displays several strong lensing features. 
  SED modeling of the lensed arc features from multicolor imaging with the
  VLT and the LBT reveals likely redshifts $\sim$2.7 for the most prominent of the
  lensed background galaxies. Mass estimates are derived for different radii 
  from the velocity dispersion of the cluster members, $M_{\rm 200} \simeq 1.8\times
  10^{14}$\,\msun, from the X-ray spectral parameters, $M_{\rm 500} \simeq 
  1.0 \times 10^{14}$\,\msun, and the largest lensing arc, $M_{\rm SL} \simeq 2.3 \times
  10^{13}$\,\msun.
  The projected spatial distribution of cluster galaxies appears to be
  elongated, and the brightest galaxy lies off center with respect to the
  X-ray emission indicating a not yet relaxed structure. XMMU\,J100750.5+125818 offers
  excellent diagnostics of the inner mass distribution of a distant cluster
  with a combination of strong and weak lensing, optical and X-ray spectroscopy.
}
\keywords{galaxies: clusters of galaxies -- X-rays: clusters of galaxies}

\maketitle

\section{Introduction}
        
A crucial prerequisite for the use of clusters of galaxies for cosmological
studies is the knowledge of the scaling relations  between cluster mass and an
observable proxy, such as X-ray luminosity, or X-ray gas temperature.  
These relations
must be calibrated at varying redshifts, using theoretical modeling and
hydrosimulations, or simply empirical relations. Clusters at low and
intermediate redshifts are bright enough in 
X-rays to measure the total gravitating mass by deprojecting their
temperature and density profiles. This is currently difficult for clusters
beyond $z\sim\! 1$ and subject to large statistical errors
\citep[e.g.][]{rosati+04}. In this high redshift regime, independent
mass estimates based on gravitational lensing (both in the weak and the strong
regimes) have long been invoked as the most effective method for calibrating
cluster masses. In particular, mass measurements based on multiple strong
lensing features are highly robust. Although such measurements are possible
only in the central region of the cluster, they are independent of model
asssumptions and can help to break the mass-sheet degeneracy of parameter-free
weak-lensing mass-reconstructions. 

To obtain a well-defined sample of clusters at high redshift, $z > 0.8$, we
have started the XDCP  \citep[XMM-Newton Distant Cluster 
Project, ][]{boehringer+05,fassbender+08,mullis+05} based
on archival XMM-Newton X-ray data. The XDCP has been very successful, so far providing
18 clusters at $z>0.8$ and 10 clusters with $z>1$ including five redshift
confirmations from other projects \citep[for selected results see
e.g.][]{fassbender+08,santos+09,rosati+09}.
        
Here we report the discovery of a distant strong lensing cluster of galaxies
at redshift $z=1.082$ which shows a giant arc and other arc-like features
interpreted as gravitationally lensed images of more distant, background objects. 
We study the properties of the lensing cluster and the lensed galaxies on the
basis of available optical and X-ray imaging and spectroscopy.

Throughout this paper we use a standard $\Lambda$CDM  cosmology with
parameters $H_0=71$\,km s$^{-1}$ Mpc$^{-1}$, $\Omega_{\rm  M}=0.27$, and
$\Omega_{\Lambda}=0.73$, which gives a scale of 8.188\,kpc arcsec$^{-1}$ 
at redshift 1.082.
All magnitudes in this paper are given in the AB system. 

\begin{figure*}[t]
\begin{minipage}{0.62\hsize}{}
\resizebox{\hsize}{!}{\includegraphics{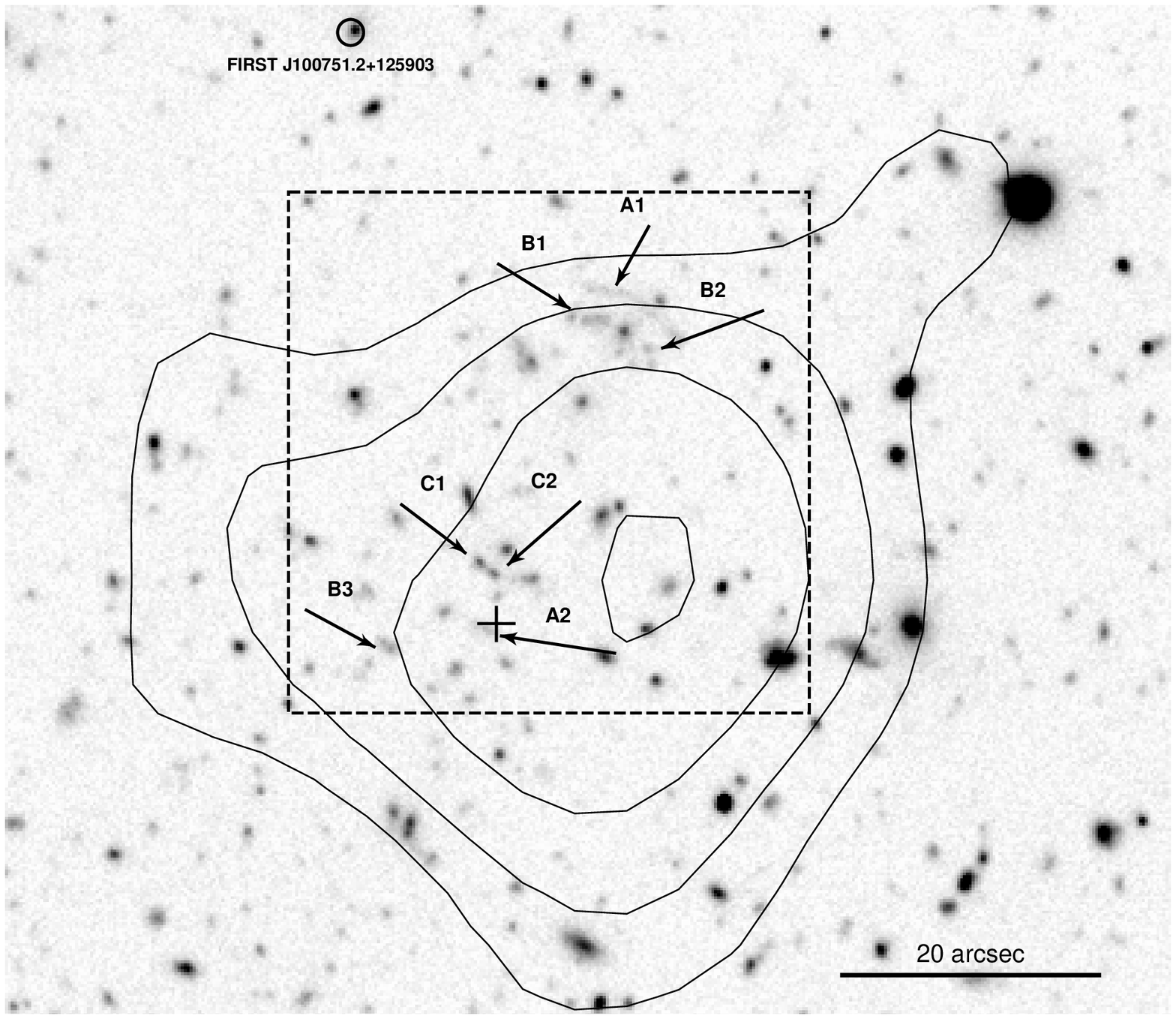}}
\end{minipage}
\hfill
\begin{minipage}[]{0.37\hsize}{}
\resizebox{\hsize}{!}{\includegraphics{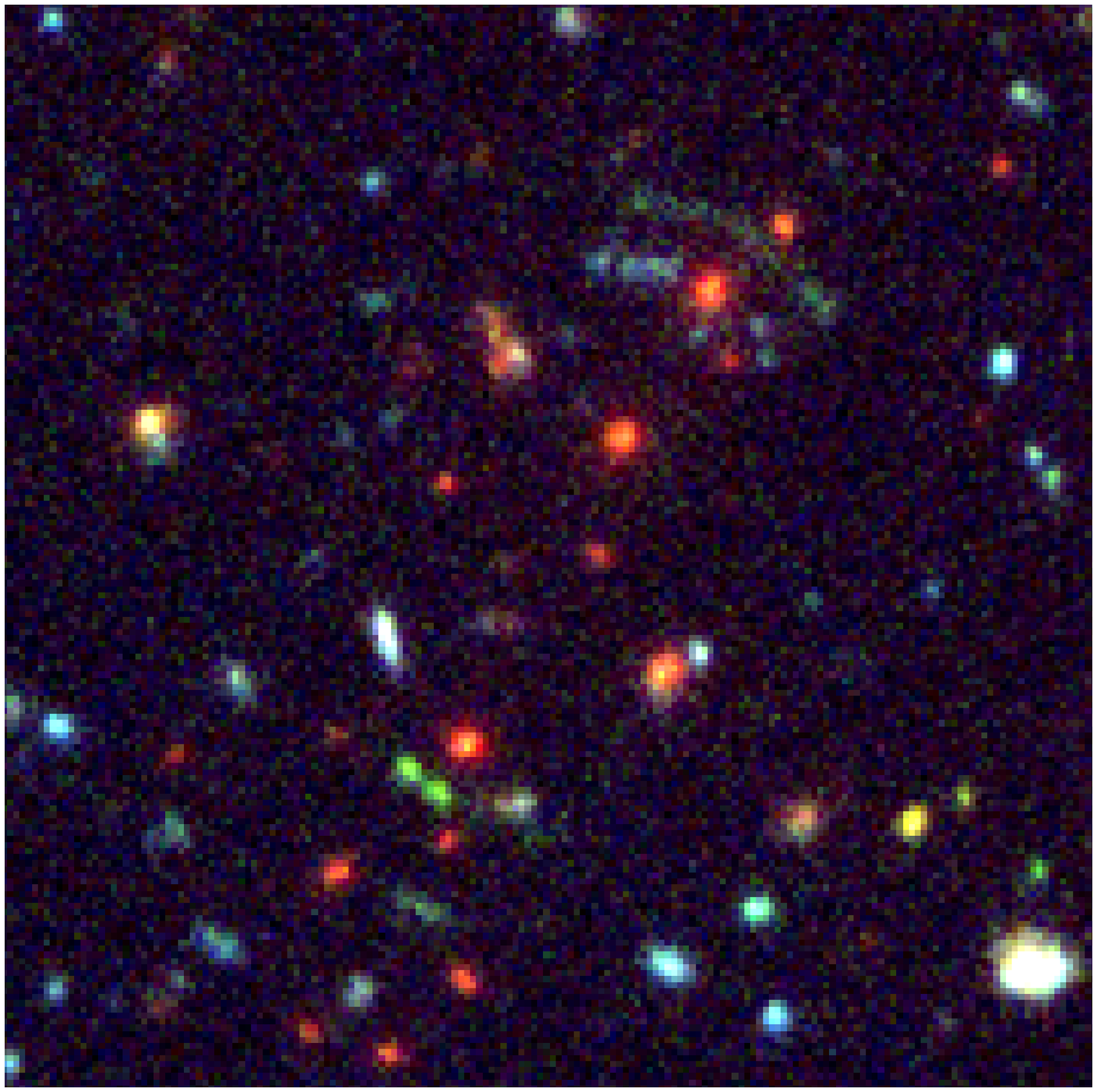}}
\caption[]{{\it} (left) UBVR image of the field around \xmmu1007\ with X-ray
  contours overlaid and tentatively identified lensing features indicated. The
  cross indicates the X-ray position and the
  square the size of the color image shown right. {\it (right)} Color
  composite of the central region of \xmmu1007\ with mean (Rz)V(UB) images
  in the RGB channels. North is top and East to the left in both
  frames. 
\label{f:opt_composit}
}
\end{minipage}
\end{figure*}

\section{Observations, analysis and results}

\subsection{XMM-Newton X-ray observations}
XMMU\,J100750.5+125818 (hereafter \xmmu1007) was found serendipitously
as an extended  X-ray source in our systematic search for distant clusters of
galaxies.
The cluster candidate was detected in
a field with nominal exposure time of \(22.2 \mathrm{ksec}\) at an off-axis
angle of \(11.7 \mathrm{'}\) (OBSID: 0140550601).   
Following \citet{pratt+arnaud03}, 
time intervals with high background were excluded by applying a two-step flare
cleaning procedure firstly in a hard energy band ($10-14$\,keV) and
subsequently in the $0.3-10$\,keV band, after which 21.4 ksec of
clean exposure time remained  for the two MOS cameras and
\(18.0\,\mathrm{ksec}\) for the PN instrument. 
        
Images and exposure maps were created in the $0.35-2.4$\,keV detection band,
which was chosen to maximize  the signal-to-noise ratio of the X-ray 
emission of  massive galaxy clusters at \(z>0.8\) compared to the background
components \citep{scharf02}. Source detection was performed
with the \texttt{SAS} task \texttt{eboxdetect}  followed by the maximum
likelihood fitting task \texttt{emldetect} for the determination of source
parameters. \xmmu1007\ was found as an extended X-ray source with low surface
brightness at position $\alpha(2000) =10^h 07^m 50\fs5, \delta(2000)=+12\degr
58' 18\farcs1$ with a total of about 200 source photons and a core radius of
\(24 \mathrm{''}\) at a significance level of DET$_{\rm ML} \sim 42$ and an
extent likelihood of EXT$_{\rm ML} \sim 24$\footnote{DET$_{\rm ML}$, EXT$_{\rm 
ML}$: detection likelihood and extent likelihood of the source, $L = - \ln P$,
where $P$ is the probability the detection (the extent) is spurious due to a
Poissonian fluctuation}.
\xmmu1007\ is not listed in the 2XMM catalog \citep{watson+09}, since 
its detection likelihood after the first step of the detection chain 
(\texttt{eboxdetect}) remained below the threshold to perform the
second step (\texttt{emldetect}). It was revealed as a cluster candidate in
this work, however, owing to the use of a different detection band
and by lowering the detection threshold to perform \texttt{emldetect}.

\subsection{Optical imaging with VLT/FORS2 and LBT/LBC}
Neither DSS nor the SDSS revealed a convincing counterpart to the unique
X-ray source \xmmu1007\footnote{We regard the NVSS radio source
   NVSS J100751+125901 at $10^h07^m51\fs3, +12\degr59'01''$ (J2000.0), marked
   in Fig.~\ref{f:opt_composit}, and 
   confirmed to be point-like by FIRST ($10^h07^h51\fs289s, +12\degr59'3\farcs34$)
   with an integrated flux of $5.59\pm0.19$\,mJy (positional offset of $48\arcsec$)
   as unrelated to the X-ray source}. 
The field containing \xmmu1007\ was thus observed with FORS2 at the VLT in
January/February 2007 through ESO filters R\_SPECIAL$+$76 and z\_GUNN$+$78 with
total integration times of 1920\,s and 960\,s, respectively. 
The combination of filters was chosen to allow an unambiguous redshift
determination up to $z\sim 0.9$ through the identification of a cluster red
sequence (CRS).  

All the imaging data of this paper were reduced with an AIP-adaptation
of the GaBoDS-pipeline described in \citet{erben+05}. It comprises
all the pre-processing steps (bias- and flatfield-, and fringe-correction)
as well as super-flat correction, background subtraction, and creation of a
final mosaic image using SWarp and ScAMP \citep{bertin06}.

The photometric calibration of the $R$-band image was achieved through
observations of \citet[][]{stetson00} 
standard fields, the photometry of the z-band image was tied to the
SDSS. The non-standard z-band cut-on filter in use at the VLT leads to an
estimated systematic calibration uncertainty of 0.05 mag.

The measured image seeing of the VLT R- and z-band images 
is $0\farcs7$  and $0\farcs56$, respectively. 
Object catalogues were generated with SExtractor
\citep{bertin+arnouts96} in double image mode.
The magnitudes of catalogued objects were corrected for
Galactic foreground extinction. From the observed drop of the number-flux 
relations with respect to a power-law expectation we estimate a
50\% completeness of the catalogues of \(R_{\mathrm{lim}} \sim 25.9\) and
\(z_{\mathrm{lim}} \sim 24.6\).     
Following \citet[][and its erratum]{mullis+05} we used fixed apertures with  
diameter \(3.5 \times  \mathrm{FWHM_{seeing}}\) to determine object colors. 
 
\begin{figure}
\resizebox{\hsize}{!}{\includegraphics[clip=]{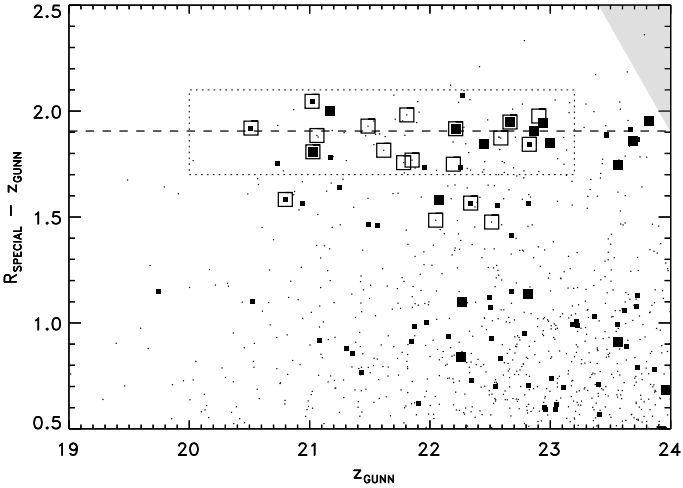}}
\caption[MD of \xmmu1007]{Color-magnitude diagram of \xmmu1007. The size of
  the symbols encodes vicinity to the X-ray center of the cluster (within
  30\arcsec, and 60\arcsec\ and beyond). Objects within the dotted box were used 
  to estimate the color of the red sequence (solid horizontal line). Objects
  framed with squares are spectroscopically confirmed cluster members. The
  shaded region top right indicates $>50$\% incompleteness.}
\label{f:cmd}
\end{figure}

A mere two-band false-color composite of the VLT imaging data proved
sufficient to locate an overdensity of red, early type galaxies at the
position of the extended X-ray emission. 
Color- and location-selected possible cluster members form a red
sequence in the color-magnitude diagram of Fig.~\ref{f:cmd}. 
The red sequence
color was estimated from the average color of all galaxies within 30\arcsec of
the X-ray position (see box in Fig.~\ref{f:cmd}), $R - z = 1.91$, which hints
to a cluster redshift beyond $\sim$0.9. 

Further imaging observations were performed with the Large
Binocular Telescope (LBT) equipped with the Large Binocular Camera (LBC) through 
Bessel B,V filters and the U$_{\rm spec}$ filter. 
Table \ref{lbt_images} lists the details of the LBT imaging observations.

The photometric zeropoints for the LBC images were derived by using
the SDSS photometry of stellar objects in the field.

A multi-color composite of the center of the field including VLT/FORS2- and
LBT/LBC-data is shown in Fig.~\ref{f:opt_composit} with X-ray contours
overlaid. Residual astrometric uncertainties in the XMM-Newton X-ray image
were removed on the basis of identified X-ray point sources in the FORS- and
LBT-images, respectively.

The cluster has its brightest galaxy (BCG) at position   
\(\alpha = 10^\mathrm{h}07^\mathrm{m}49.9^\mathrm{s}\) and 
\(\delta = +12^\mathrm{\circ}58^\mathrm{'}40^\mathrm{''}\)
between structures B1 and B2 (see Fig.~\ref{f:opt_composit}), i.e.~away from the
apparent center of the galaxy distribution by about $20''$ and away from the
X-ray position. 
The BCG has an apparent magnitude $z_\mathrm{BCG} = 20.51 \pm 0.06$, an
osberved color $(R-z)_\mathrm{BCG} = 1.92 \pm 0.08$ and an absolute magnitude
$M_{\rm BCG, z} = -23.8$. 

The spatial distribution of galaxies that form the red sequence of
Fig.~\ref{f:cmd} appears 
stretched in a band from SSE to NNW with a centroid about $10''$
away from the centroid of the cluster X-ray emission. However, the best-fit 
X-ray position (Maximum Likelihood fit of an assumed $\beta$-profile folded
with the PSF of the X-ray telescope) lies well within the galaxy assembly (see 
Fig.~\ref{f:opt_composit}). 

\subsection{Optical spectroscopy}
Multi-object spectroscopy of \xmmu1007\ was performed with VLT/FORS2 using the
MXU option in  December 2007 and January 2008. One mask was
prepared targeting 37 candidate objects selected as possible cluster members
based on color and stellarity index. Individual exposures of 1308\,s were combined
to yield a total integration time per object of 2.9\,h. Spectra  were obtained with
grism 300I; they cover the wavelength range $6000-11000$\,\AA\ with a scale of
3.2\,\AA/pixel at the chosen $2\times2$\,binning. 

We obtained 32 classifiable spectra (among them two late-type stars) and
measured the redshifts of galaxies by fitting double Gaussians with fixed
wavelength ratio and same width to the Ca H\&K lines.
The measured redshifts together with coordinates, brightness, and color of
the 32 objects are listed in Table~\ref{t:cluster_members} and marked on
Fig.~\ref{f:chart_spectra} (appendix).

We found 19 concordant redshifts between $1.075$ and $1.088$ with a weighted mean
of 1.08103 and a median of 1.08207. We assume a cluster redshift of 1.082 in
the following. The redshifts in this interval are Gaussian distributed, a
one-sample KS-test reveals a probability of rejecting the null hypothesis of
1.3\%. The distribution of all redshifts is shown in Fig.~\ref{f:zhist}.
Following \citet{danese+80}  we calculate a line-of-sight velocity
dispersion of $\sigma_{\rm p} = 572$\,\kmps with a 90\% confidence range between
437\,\kmps and 780\,\kmps. 
        
Although the cluster does not appear to be relaxed, we may estimate a mass on
the virial assumption. Using $R_{200} =\sqrt{3}/10 \sigma/H(z), M_{200} =
4/3\pi R_{200}^3 \times 200\rho_c$ \citep{carlberg+97a}, we obtain $R_{200} =
790^{+280}_{-185}$\,kpc, and $M_{200} = 1.8^{+2.8}_{-1.0} \times
10^{14}$\,\msun\ where the given errors correspond to the 90\% confidence
interval of the velocity dispersion. Using the dark matter halo virial scaling
relation by \citet{evrard+08} instead, one obtains $M_{200} = 1.2^{+1.8}_{-0.7} \times
10^{14}$\,\msun. 

\subsection{X-ray spectroscopy}
X-ray spectra were extracted from the calibrated photon event lists of all
three EPIC cameras onboard XMM-Newton. An aperture with 60\arcsec\
radius ($\sim$500\,kpc at the cluster distance) was used to extract source and
background photons.  The X-ray spectrum of \xmmu1007\ 
contains $\sim$200 net photons in the full XMM-band after background subtraction. 

An attempt was made to constrain the plasma temperature with a thin thermal
plasma model \citep[a MEKAL model in XSPEC terms,][]{mewe+85}. 
The column density of absorbing matter was fixed to its galactic value,
$N_\mathrm{H} = 3.7 \times 10^{20}$\,cm$^{-2}$, the metal abundance to
$Z = 0.3 Z_\mathrm{\sun}$, and the redshift to $z=1.082$.

Despite fixing these parameters, the X-ray temperature could only be roughly
constrained. The fit shown in Fig.~\ref{f:xray_spectra} converges at a
temperature of $T = 5.7$\,keV ($kT > 2.05$\,keV with 90\%\,confidence) 
implying a bolometric X-ray luminosity of 
$L_\mathrm{X} = 1.3 \times 10^{44}$\,erg\,s$^{-1}$
($L_{\rm X} > 0.8\times 10^{44}$\,erg\,s$^{-1}$ at 90\% confidence,
$L_\mathrm{X}(\mathrm{0.5-2.0\,keV}) = 4 \times 10^{43}$\,erg\,s$^{-1}$). 

Following \citet[][]{pratt+09} we estimate the cluster mass using
the luminosity-mass scaling relation. We use their BCES orthogonal fit to the
Malmquist-bias corrected L-M relation, assume self-similar evolution,
$h(z)^{-7/3}$, and obtain $M_{500} = 1.0\times 10^{14}$\,\msun\ 
(assuming our aperture covers $R_{500}$). 
If we instead use the temperature-mass relation of \citet{vikhlinin+09}
we find $M_{500}=2.1\times10^{14}$\,\msun\ with a lower limit of
$M_{500}=4.3\times10^{13}$\,\msun.

\subsection{Lensing properties}
Our follow-up imaging observations confirmed the existence of lensing arcs and
further lensing features in the image. We used the imaging data in the
five passbands (UBVRz) from the LBT and the VLT to calculate photometric
redshifts of the lensed background objects. Due to the distorted morphology of
the lensing arcs we manually defined apertures matching the shape of each
feature. The fluxes of the objects were then extracted using the same aperture
in each image and converted to AB magnitudes.

\begin{table}[thb]
\caption[LBT imaging]{Photometric redshifts of lensing features. All redshifts
  were forced to be larger than the cluster redshift.}
\centering
\begin{tabular}{llll}  \hline\hline
Image  &  $V_{\rm AB}$     & $z_{\rm phot}$ &  $z$ range (90\%) \\ \hline
 A1    &   24.3            &   2.72        &   2.63 - 2.82     \\
 A2    &   25.1            &   2.63        &   2.54 - 2.73     \\
 B1    &   24.5            &   1.39        &   1.35 - 1.52    \\
 B2    &   25.7            &   1.63        &   1.08 - 1.74   \\
 B3    &   24.7            &   1.94        &   1.70 - 2.12    \\
 C1    &   24.3            &   3.36        &   2.98 - 3.54    \\
 C2    &   23.6            &   3.20        &   2.70 - 3.40    \\
\hline
\end{tabular}
\label{zphot}
\end{table}
    
We used the publicly available {\tt hyperz} code \citep{bolzonella+00}
to compare the spectral energy distributions of the lensing features with the
synthetic galaxy  SEDs from \citet{bruzualcharlot93}. The parameters for
the construction of the SED data cube were galaxy class, star forming age,
internal reddening ,  and redshift. For the lens features we allowed only
redshifts beyond the redshift of the cluster ($z=1.082$) and  combinations of age
and redshift consistent with our adopted cosmological parameters. 

Table~\ref{zphot} shows the results of the SED fitting for the image
components as indicated in Fig.~\ref{f:opt_composit}. Column (2) gives the best
fitting photometric redshift and column (3) the 90\% confidence limits of
the redshift within the best fitting SED template. 

The resulting photometric redshifts confirm our tentative identification of
multiple lensed components. 
Components A1/A2 and C1/C2 are likely to be images of the same background
objects at $z\sim 2.7$ and $z\sim{3}$, respectively.  
The situation is less clear for components B1, B2, and B3, where the
uncertainties in the photometric redshifts are large. 

We estimate a lensing mass assuming a circularly symmetric lens (a special
case is the singular isothermal sphere -- SIS). 
The projected mass inside a tangential arc then becomes $M(\theta) = \Sigma_{\rm cr} \pi
(D_{d}\theta)^2 \simeq 1.1 \times 10^{14} \left(
  \frac{\theta}{30''}\right) \left(
  \frac{D}{1\mathrm{Gpc}}\right)$\,\msun. The effective distance $D$ becomes
$D = \frac{D_{\rm d} D_{\rm ds}}{D_{\rm s}} = 724$\,Mpc for $z_{\rm d} = 1.082$
and $z_{\rm s} = 2.7$. 
 
Difficulties arise from
the faintness of the lensing features and from the badly determined cluster
center. A trace of the feature A1 implies an Einstein radius of $\theta \sim
8''-9''$ and a corresponding mass of $(2.3\pm0.4)\times10^{13}$\,\msun\
(a 15\% uncertainty in radius is assumed).  
If one uses uses the distance between A1 and the X-ray center ($21\farcs5$)
instead, the mass inside this ring becomes $5.7\times10^{13}$\,\msun.  

\section{Discussion and conclusions}

We have presented results of an initial study of the X-ray selected cluster
\xmmu1007. We determine a redshift of $z=1.082$ based on 19 spectroscopic
confirmed members. As most prominent property, the cluster shows several
strong lensing features. It is 
an optically rich cluster; the absence of a dominant BCG and the elongated
distribution of member galaxies suggest a not yet relaxed
structure. The bolometric X-ray luminosity is  \(L_\mathrm{X} \approx 1.3
\times 10^{44} \mathrm{ergs/s}\). 

\begin{figure}[t]
\resizebox{\hsize}{!}{\includegraphics[clip=]{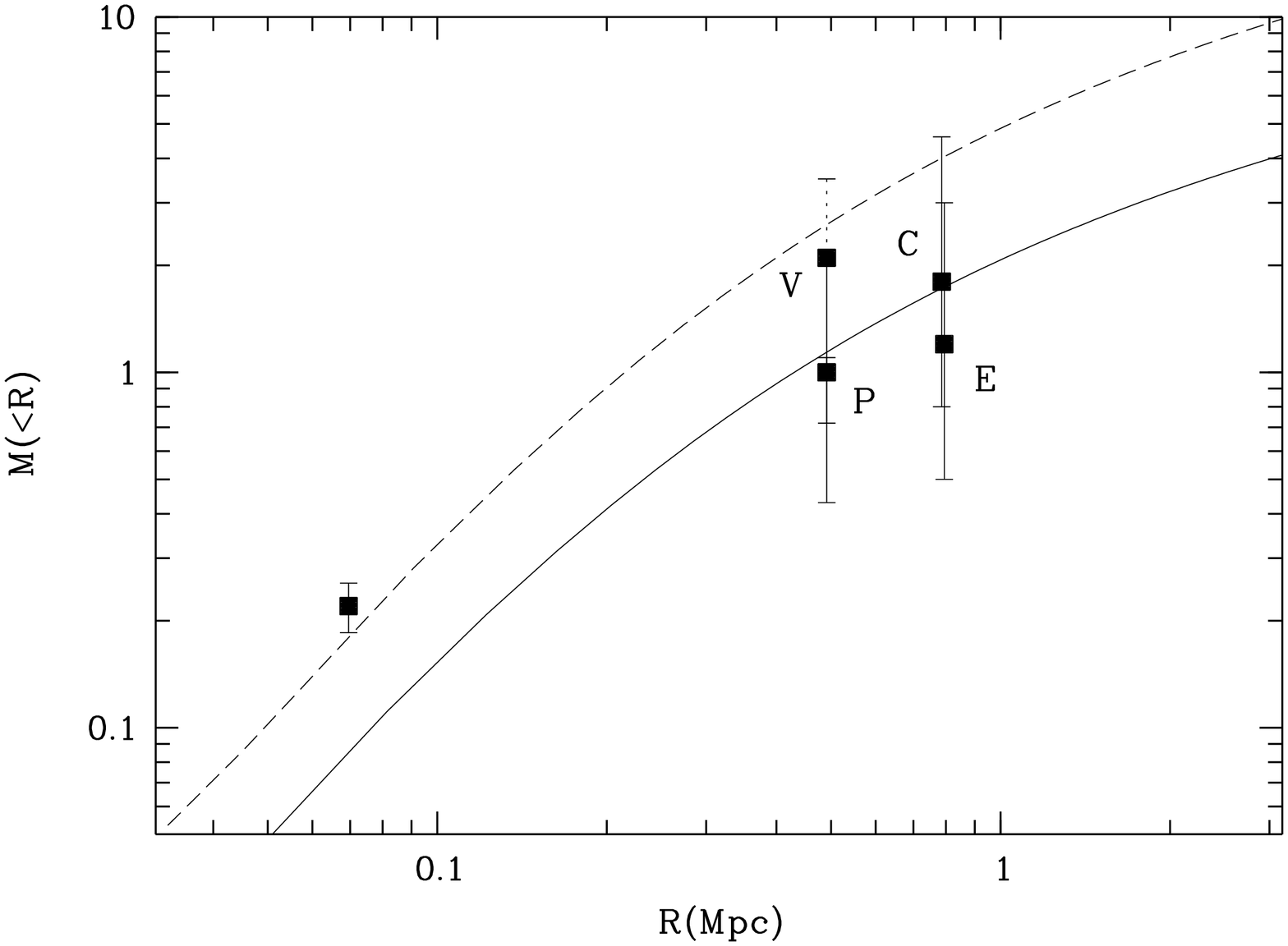}}
\caption[]{Comparison of different mass estimates for \xmmu1007\ in units of
  $10^{14}$\,\msun. Labels indicate results based an scaling relations by
  \citet{vikhlinin+09,pratt+09,carlberg+97a,evrard+08}. The dashed and solid
  lines indicate universal density profiles \citep{nfw97} with concentration
  parameters of 3.86 and 3.45 for virial masses 1.8 and $4.6\times
  10^{14}$\,\msun\ at an assumed $R_{\rm vir} = 780$\,kpc, respectively \citep{bullock+01}.
}
\label{f:nfw}
\end{figure}

Estimates of the cluster mass were obtained via strong lensing, X-ray
spectroscopy and the velocity dispersion of member galaxies; the values
obtained at different radii are summarized in Fig.~\ref{f:nfw}. 
One obtains overlapping error bars for the mass at $\sim$$R_{200}$ ($\sim$virial
radius) and at $\sim$$R_{500}$ ($\sim$X-ray aperture) due to insufficient data 
and uncertainties in the scaling relations. 
If one takes $M_{\rm vir} \simeq M_{200} = 1.8\times10^{14}$\,\msun\ at face
value, our strong lensing result seems to be discrepant with an
NFW-profile. All data can be made compliant at 
a mass of about $4\times10^{14}$\,\msun, higher than but not excluded by the
velocity dispersion and the X-ray temperature, whereas a weak-lensing mass is key to
fix the halo profile at large radii, a deep X-ray observation is necessary to
determine the X-ray morphology and the gas temperature. The modelling of
the strong lensing system is important to probe the inner density profile. 

Interestingly, 
in a study of a complete sample of 12 MACS clusters \citet{zitrin+10} find the
observed Einstein radii to be larger and hence the central 
density of cluster higher than predicted by simulations and interpret their
finding as a challenge to cluster formation in a $\Lambda$CDM model. They invoke
the possibility that the formation of clusters started at earlier epochs than
currently assumed, leading to higher central Dark Matter concentrations. 
Also, \citet{jee+09}
suggest this scenario as a possible explanation for the discovery of
unexpectedly massive clusters at $z>\sim$1 in the moderate survey volume 
probed by the XMM-Newton serendipitous surveys \citep[like
e.g.~XMMU\,J2235-2557 and  2XMM\,J083026+524133,][]{mullis+05,lamer+08,rosati+09}.  
However, in a recent study of the strong lensing clusters in the {\sc Mare Nostrum
Universe}, Meneghetti et al.~(2010, subm.~to A\&A) find the 
concentration and the X-ray luminosity to be biased high, and an excess of
kinetic energy within the virial radius among the strong lensing
clusters. \xmmu1007\ is a highly suited  target to confront theory and
observation.  

The elongated distribution of the cluster member galaxies
may point to the principal direction of merging or accretion
of \xmmu1007\ \citep{dubinski98}.
The merging hypothesis is consistent with the system being a strong lens
\citep[see e.g.~the chain distribution of the brightest member galaxies
in the strong-lensing main component of the Bullet Cluster,][]{bradac+06}.
On the other hand, the chain distribution of the bright member galaxies
of \xmmu1007\ may suggest that the infall of these galaxies
happens along filaments with small impact parameters,
so that dynamical friction is particularly efficient in dragging them
to the cluster center and originate a dominant BCG within a short time
\citep{donghia+05}.
In analogy with the formation of fossil galaxy groups
studied by the latter authors, the small impact parameters
of the filaments may lead to an early assembly of the gas in the center,
thus to an early start of its cooling.
The resulting system may exhibit an enhanced X-ray luminosity
with respect to the optical one.
This efficient, early formation may offer an alternative explanation
to merging for the reason why particularly X-ray luminous clusters
with chain distribution of their bright members are detected
at high redshifts.

\begin{acknowledgements}
We thank J.~Wambsganss for providing software code describing the lensing
geometry. We thank S. Schindler and W. Kausch for an early analysis of the
lensing features. HQ thanks the FONDAP Centro de Astrofisica for partial
support. This work was supported by the DFG under grants Schw
536/24-2 and BO 702/16-3, and the German DLR under grant 50 QR 0802.

We thank our referee, Florence Durret, for constructive criticism.

Based on data acquired using the Large Binocular Telescope (LBT). The
LBT is an international collaboration among institutions in the United
States, Italy, and Germany. LBT Corporation partners are: The University
of Arizona on behalf of the Arizona university system; Istituto
Nazionale di Astrofisica, Italy; LBT Beteiligungsgesellschaft, Germany,
representing the Max-Planck Society, the Astrophysical Institute
Potsdam, and Heidelberg University; The Ohio State University, and the
Research Corporation, on behalf of The University of Notre Dame,
University of Minnesota, and University of Virginia.
\end{acknowledgements}

\bibliographystyle{aa} 
\bibliography{14430}

\clearpage 
\appendix

\section{Supplementary electronic material}
This appendix gives further detailed information about optical observations,
the X-ray spectrum, the redshift distribution and the brightness of
spectroscopically classified objects in the field of \xmmu1007. 
\clearpage

\begin{table}[bht]
\caption[LBT imaging]{LBT image parameters}                                
\centering
\begin{tabular}{lllll} \hline\hline
filter   &date          & exposure & seeing & zeropoint  \\  
\hline
U$_{\rm spec}$   & 31-Dec-2008  & 1200 s    & $0\farcs65$ & 26.73 \\
B$_{\rm Bessel}$ & 31-Dec-2008  & 1070 s    & $0\farcs75$ & 27.41 \\
V$_{\rm Bessel}$ & 31-Dec-2008  & 1200 s    & $0\farcs78$ & 28.01 \\
\hline
\end{tabular}                        
\label{lbt_images}
\end{table}

\begin{figure}
\resizebox{\hsize}{!}{\includegraphics[clip=]{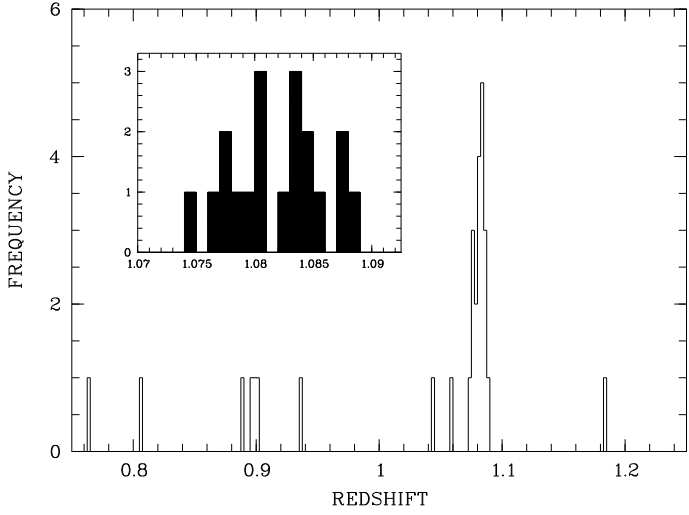}}
\caption[]{Redshift distribution of galaxies in the field of \xmmu1007\
observed with VLT-FORS2-MXU (see Table~\ref{t:cluster_members}). 
The inset is a blow-up comprising the redshift range covered by likely cluster
members.}
\label{f:zhist}
\end{figure}
        
\begin{figure}
\resizebox{\hsize}{!}{\includegraphics[angle=-90,clip=]
  {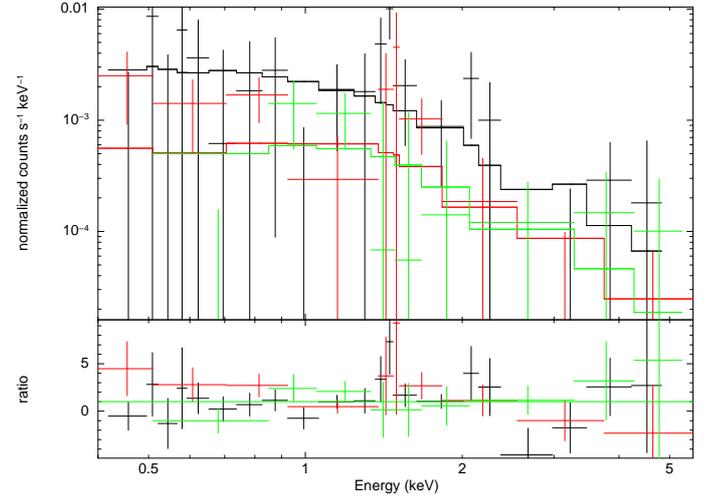}}
\caption[X-ray spectrum of \xmmu1007]
{XMM-Newton X-ray spectrum of \xmmu1007 (EPIC-pn in black, EPIC-MOS in
red and green). A thin thermal plasma model (the so-called MEKAL-model in
the X-ray spectral fitting package {\em XSPEC})
was fitted to the data. Fit optimisation was achieved by using the 
Cash-statistic with one count per bin \citep{krumpe+08}. It does not,
however, provide a goodness-of-fit statistic. The fit suggests a
cluster temperature of $T = 5.7$\,keV ($kT > 2.05$\,keV with 90\%
confidence). The temperature is not constrained towards high values. The
bolometric X-ray luminosity of this model is $L_\mathrm{X} = 1.3 \times 
10^{44}$\,ergs\,s$^{-1}$.} 
\label{f:xray_spectra}
\end{figure}
        
\begin{table*}[hb]
\caption[Results from the spectroscopy campaign with respect to 
XMMU1007; member and non-member galaxies]{Spectroscopically confirmed cluster
  member galaxies (upper part) and non-member galaxies targeted by VLT-MOS
  spectroscopy sorted by increasing z-band magnitude.  
  The 2nd to last column lists the positional offset with respect to the
  assumed cluster center at the nominal X-ray position
  (R.A. $151.9604^{\circ}$, Decl. $12.9717^{\circ}$). 
  A colon indicates a less secure measurement, which may be subject
  to future adjustment.} 
\label{t:cluster_members} 
\centering
\begin{tabular}{l|c|c|c|c|c|c|c|l}
\hline \hline
ID & R.A. J2000 & Decl. J2000 &  z & R--z & redshift $z$ & $\delta\, z$ & Distance \\
& (deg) & (deg) & (mag) & (mag) & & & (arcsec) &\\
\hline
  1-10  & 151.95762  & 12.97789  &  20.51  & 1.92  & 1.08423  & 0.00012 &  24.1    \\ 
  1-5   & 151.96234  & 12.96729  &  20.80  & 1.59  & 1.08086  & 0.00012 &  17.8    \\                        
  1-9   & 151.95854  & 12.97640  &  21.02  & 2.05  & 1.08709  & 0.00021 &  17.8    \\                        
  1-8   & 151.95809  & 12.97400  &  21.03  & 1.81  & 1.07712  & 0.00078 &  11.3    \\                        
  1-19  & 151.91809  & 13.00307  &  21.06  & 1.89  & 1.07834  & 0.00016 & 187.0    \\                        
  1-20  & 151.92177  & 13.00492  &  21.49  & 1.93  & 1.08340  & 0.00014 & 181.3    \\                        
  2-14  & 151.96438  & 12.95042  &  21.62  & 1.82  & 1.08715  & 0.00008 &  77.9    \\                        
  2-5   & 151.91762  & 12.92344  &  21.78  & 1.76  & 1.08309  & 0.00011 & 229.7    \\                        
  1-14  & 151.95837  & 12.98867  &  21.81  & 1.98  & 1.07465  & 0.00009 &  61.5    \\                        
  1-21  & 151.92233  & 13.00835  &  21.85  & 1.77  & 1.08207  & 0.00010 & 188.3    \\                        
  1-16  & 151.92898  & 12.99330  &  22.05  & 1.48  & 1.07995  & 0.00009 & 135.0    \\                        
  2-9   & 151.92196  & 12.93457  &  22.19  & 1.75  & 1.07749  & 0.00009 & 189.9    \\                        
  1-7   & 151.96153  & 12.97194  &  22.21  & 1.92  & 1.08463  & 0.00013 &   4.6    \\                        
  1-3   & 151.96457  & 12.96373  &  22.34  & 1.57  & 1.08069  & 0.00022 &  32.6    \\                        
  2-1   & 151.91657  & 12.91210  &  22.51  & 1.48  & 1.08085: & 0.00066 & 263.7    \\ 
  1-15  & 151.94809  & 12.99093  &  22.59  & 1.88  & 1.08370  & 0.00017 &  81.5    \\                        
  1-6   & 151.96098  & 12.97008  &  22.67  & 1.95  & 1.07673  & 0.00017 &   6.5    \\                        
  1-2   & 151.956    & 12.96175  &  22.83  & 1.85  & 1.08538: & 0.00033 &  39.1    \\ 
  1-13  & 151.95492  & 12.98622  &  22.90  & 1.98  & 1.08834: & 0.00020 &  55.6    \\ 
\hline
  1-23  & 151.97453  & 13.01627  &  18.74  & 3.16  & 0.0      & 0.0      & 167.9& late-type star \\ 
  1-11  & 151.94881  & 12.98072  &  20.06  & 2.14  & 0.0      & 0.0      & 52.0 & late-type star \\ 
  1-12  & 151.96350  & 12.98431  &  20.74  & 1.75  & 0.89789  & 0.00022  & 46.7 & [OII] \\ 
  1-1   & 151.94980  & 12.95957  &  20.94  & 1.57  & 1.05870  & 0.00021  & 57.4  \\ 
  2-11  & 151.94960  & 12.94159  &  21.25  & 1.47  & 0.88879  & 0.00015  & 114.8 & [OII]\\ 
  1-18  & 151.98341  & 12.99842  &  21.87  & 1.80  & 1.18291  & 0.00008  & 125.6 \\                            
  2-3   & 151.95777  & 12.91942  &  22.19  & 1.86  & 1.03811: & 0.00011  & 188.4 \\ 
  1-22  & 151.96277  & 13.01301  &  22.21  & 1.35  & 0.80748  & 0.00013  & 148.9 & [OII]\\ 
  2-2   & 151.96559  & 12.91600  &  22.24  & 1.42  & 0.89511  & 0.00021  & 201.3 \\ 
  2-7   & 151.93620  & 12.92835  &  22.43  & 1.62  & 0.90020  & 0.00027  & 177.6 \\ 
  2-8   & 151.97170  & 12.93165  &  22.58  & 1.82  & 1.04382: & 0.00020  & 149.5 \\ 
  2-6   & 151.91717  & 12.92623  &  22.68  & 1.36  & 0.76364  & 0.00014  & 223.1 \\ 
  2-10  & 151.96192  & 12.93803  &  23.73: & 0.54  & 0.93729  & 0.00019  & 121.3 \\ 
\hline
\end{tabular}
\end{table*}

\begin{figure*}
\resizebox{\hsize}{!}{\includegraphics[clip=]{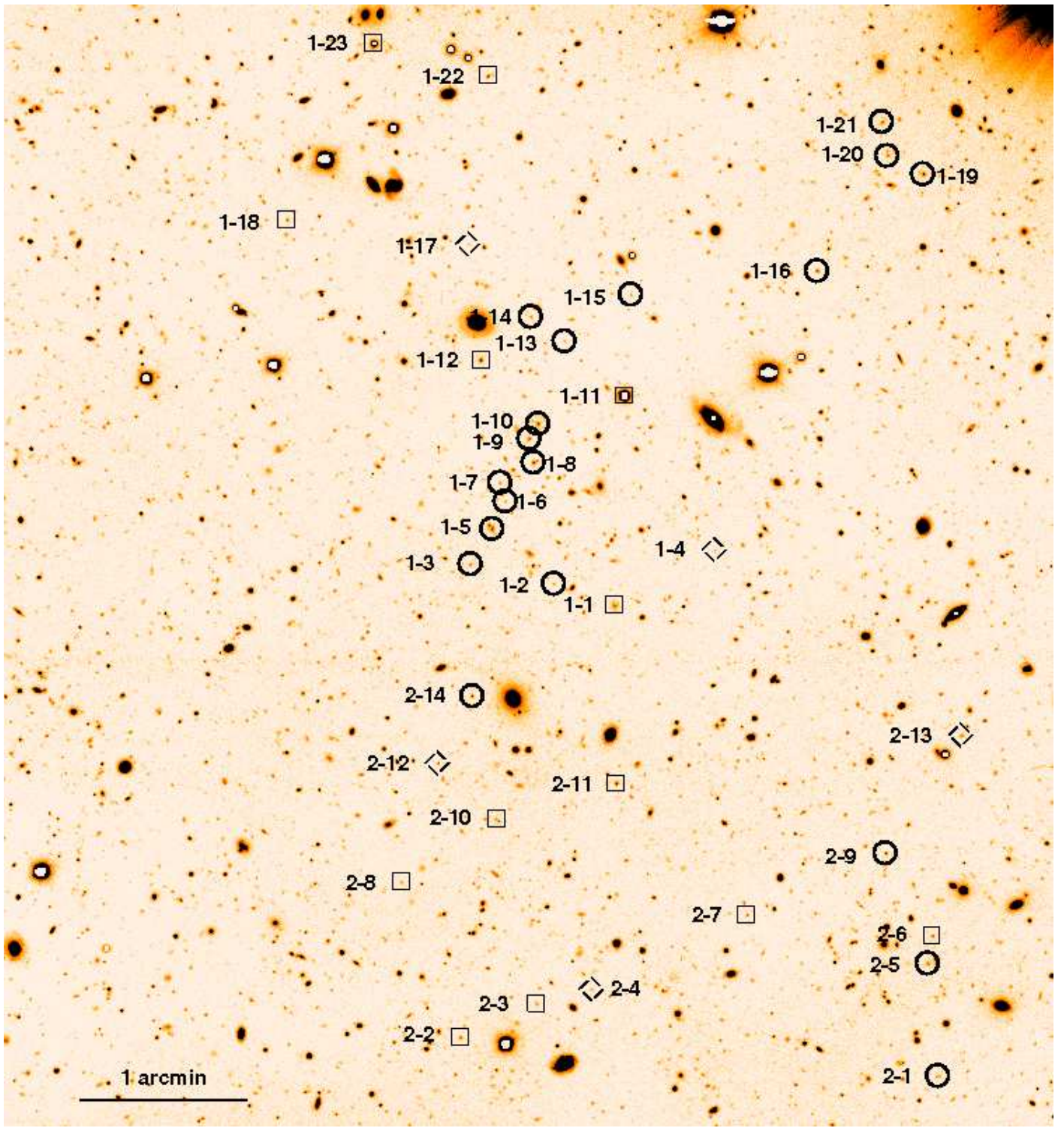}} 
\caption[]{VLT/FORS2 R-band image with slit positions identified (North is
  top, East to the left). Labels are the same as in the first column of
  Table~\ref{t:cluster_members}.
  Likely cluster members are encircled, unrelated classified 
  objects (lower section of Table~\ref{t:cluster_members}) are framed with
  squares, objects that were positioned on slits but could not yet be
  classified are framed with rhombs.} 
\label{f:chart_spectra}
\end{figure*}
                
\end{document}